\documentclass[aps,pre,twocolumn,showpacs]{revtex4-2}

\usepackage{amssymb,amsmath}
\usepackage{bm}
\usepackage{color}

\usepackage{graphicx}

\begin{document}
	
\widetext
\textit {This is the version of the article before peer review or editing, as submitted by an author to EPL (Europhysics Letters). IOP Publishing Ltd is not responsible for any errors or omissions in this version of the manuscript or any version derived from it. The Version of Record is available online at \url{https://doi.org/10.1209/0295-5075/134/68002}. }

\title[Rotating spherical particle in a continuous viscoelastic medium]{Rotating spherical particle in a continuous viscoelastic medium\\
		--- a microrheological example situation}\label{title}
	
\author{S. K. Richter}
\author{C. D. Deters} 
\author{A. M. Menzel}
\email[E-mail: ]{a.menzel@ovgu.de}
\affiliation{Institut f\"ur Physik, Otto-von-Guericke-Universit\"at Magdeburg, Universit\"atsplatz 2, 39106 Magdeburg, Germany.}



\begin{abstract}
	Using analytical calculations, we characterize the rotational behavior of a rigid spherical particle when subject to a net external torque in a continuous viscoelastic environment. On long time scales, the embedding medium can either feature a net terminal flow, like a fluid, or damped reversible dynamics, like an elastic solid. The coupling of the sphere to its environment together with the therein induced deformations and flows are taken into account explicitly. In reality, using magnetically anisotropic particles, the torque can, for instance, be applied via magnetic fields. We calculate corresponding response functions. This connects our study to evaluations of microrheological investigations.
\end{abstract}

\vspace{0.7cm}
\maketitle

\section{Introduction\label{intro}}

There are several possible reasons why sometimes macroscopic rheological measurements to determine the static or dynamic mechanical properties of viscoelastic substances may be out of reach or not appropriate. For instance, it might not be possible to generate appropriately shaped samples to fit into the rheometer, or the necessary macroscopic amount of material may not be available. Moreover, samples may very sensitively and in a non-predictable way depend on the circumstances during their fabrication. In these cases, one might wish to perform rheological measurements on the same sample that is later investigated or practically used in other settings and prepared accordingly. This, for example, applies to viscoelastic gel-like substances prepared close to the critical crosslinking density \cite{huang2016buckling,puljiz2016forces}. 

Microrheological approaches may provide a solution to many of such problems. In this framework, one tracks the configurational changes of nano- to microscopic discrete particles embedded in the viscoelastic media under study \cite{mackintosh1999microrheology,crocker2000two, levine2000one, waigh2005microrheology,wilhelm2008out,wirtz2009particle, squires2010fluid,puertas2014microrheology,paul2019active}. In passive microrheology, the mechanical parameters of the viscoelastic environment can be determined by recording the configurational changes induced by thermal fluctuations. The particles need to be small enough for this method to work in a reasonable way, typically in the submicrometer range. Conversely, in active microrheology configurational changes are induced by external stimuli. From the magnitude and time sequence of the response, the rheological properties of the surroundings of the probe particles are extracted. 

A convenient approach of imposing in a noninvasive way the external stimulus to which the probe particles respond is provided by magnetic microrheology \cite{mackintosh1999microrheology,bausch1999measurement, raikher2006magnetic,wilhelm2008out, roeder2012shear,roeben2014magnetic,hess2020scale}. There, the probe particles react to magnetic fields applied from outside. This method becomes even more suitable when the system by itself already consists of magnetic particles embedded in a viscoelastic medium, and the dynamics of these magnetic constituents can be exploited for microrheological purposes. Such situations are provided by magnetic gels or elastomers \cite{filipcsei2007magnetic,odenbach2016microstructure, weeber2018polymer}. Possible applications of these systems are discussed, for instance, in the framework of soft actuators \cite{bose2012soft,hines2017soft,fischer2020towards} or soft materials of tunable mechanical stiffness \cite{jolly1996magnetoviscoelastic,schumann2017situ}.

In the present work, we derive for a linearly responding viscoelastic medium the rotational response of embedded, mutually noninteracting, spherical rigid particles under the influence of an imposed external dynamic torque. In contrast to several other approaches, we explicitly include into our calculations the distortions and flows induced in the viscoelastic environment. Our theory allows by adjusting one parameter to interpolate between solid-like media of perfectly reversible, elastic response as one limiting case and viscous fluids of vanishing elastic memory as the opposite limit. A continuous range of viscoelastic systems is found between these two limits. Linear response functions for the particle rotations under the influence of the external torques are determined. We assume the particles to be large enough so that thermal noise may be neglected. As an important example situation, we consider magnetically hard particles exposed to oscillating external magnetic fields. Our derived expressions for the resulting magnetic susceptibility may find its application in evaluations of corresponding microrheological measurements. 

\vfill
\section{Viscoelastic medium}

We adopt an Eulerian point of view when we describe the distortions and flows of the embedding viscoelastic medium as well as the dynamics of the enclosed spherical particle \cite{temmen2000convective, puljiz2019memory}. That is, the dynamic state of the medium at a certain point in time $t$ is described by a flow field $\mathbf{v}(\mathbf{r},t)$. In addition, the elastic character of the medium is reflected by a displacement field $\mathbf{u}(\mathbf{r},t)$. The latter quantifies the \textit{remaining} reversible elastic displacement that a material element presently located at position $\mathbf{r}$ has experienced to end up at this position. Or, the other way around, if the medium relaxed from its current state at time $t$ to an unstressed state, the material element presently located at position $\mathbf{r}$ would displace to $\mathbf{r}-\mathbf{u}(\mathbf{r},t)$. Thus, $\mathbf{u}(\mathbf{r},t)$ represents a type of \textit{memory} field. Over time, this memory decays, unless the medium is perfectly elastic. We confine ourselves to linearized overdamped dynamics for incompressible systems, that is $\bm{\nabla}\cdot\mathbf{v}(\mathbf{r},t)=0$ and $\bm{\nabla}\cdot\mathbf{u}(\mathbf{r},t)=0$.
 
In the appendix of ref.~\cite{puljiz2019memory}, it was shown that the following dynamic equation holds under these circumstances:
\begin{equation}\label{eq_uv}
\mu\nabla^2\mathbf{u}(\mathbf{r},t)+\eta\nabla^2\mathbf{v}(\mathbf{r},t)=\bm{\nabla}p(\mathbf{r},t)-\mathbf{f}_{\mathrm{b}}(\mathbf{r},t). 
\end{equation}
\noindent In the elastic case, $\mu$ plays the role of an elastic shear modulus, $\eta$ in the hydrodynamic limit corresponds to the dynamic viscosity, $p(\mathbf{r},t)$ represents the pressure field, and $\mathbf{f}_{\mathrm{b}}(\mathbf{r},t)$ sets the bulk force density acting on the medium. A basic relaxational behavior is assumed for the memory displacement field $\mathbf{u}(\mathbf{r},t)$ in addition to its dynamic driving by the flow $\mathbf{v}(\mathbf{r},t)$,
\begin{equation}\label{eq_gamma}
\mathbf{\dot{u}}(\mathbf{r},t)=\mathbf{v}(\mathbf{r},t)-\gamma\mathbf{u}(\mathbf{r},t). 
\end{equation}
\noindent Here, $\gamma$ is a relaxation parameter that sets the forgetfulness of the medium concerning the previous locations of its material elements. For $\gamma=0$, the medium is perfectly elastic. Then $\mathbf{u}(\mathbf{r},t)$ corresponds to the usual elastic displacement field, and all deformations are completely reversible. In this case, solid-like materials are characterized, the dynamics of which is damped as set by the parameter $\eta$. Conversely, $\gamma\rightarrow\infty$ characterizes the viscous hydrodynamic limit, lacking any elastic contribution. 

After solving eq.~(\ref{eq_gamma}) for $\mathbf{v}(\mathbf{r},t)$ and inserting it into eq.~(\ref{eq_uv}), we find the basic underlying dynamic equation for $\mathbf{u}(\mathbf{r},t)$ \cite{puljiz2019memory}: 
\begin{equation}\label{eq_u}
(\mu+\gamma\eta)\nabla^2\mathbf{u}(\mathbf{r},t)+\eta\nabla^2\mathbf{\dot{u}}(\mathbf{r},t)=\bm{\nabla}p(\mathbf{r},t)-\mathbf{f}_{\mathrm{b}}(\mathbf{r},t). 
\end{equation}
\noindent Imposing the force impact $\mathbf{F}$ at position $\mathbf{R}_0$ and time $t_0$ by setting $\mathbf{f}_{\mathrm{b}}(\mathbf{r},t)=\mathbf{F}\delta(\mathbf{r}-\mathbf{R}_0)\delta(t-t_0)$, eq.~(\ref{eq_u}) is solved by deriving the corresponding Green's function \cite{puljiz2019memory}. It quantifies the resulting memory displacement field $\mathbf{u}(\mathbf{r},t)=\underline{\mathbf{G}}(\mathbf{r}-\mathbf{R}_0,t-t_0)\cdot\mathbf{F}$. Spatial and temporal parts factorize, $\underline{\mathbf{G}}(\mathbf{r},t)=\underline{\mathbf{G}}(\mathbf{r})G(t)$, where
\begin{equation}\label{eq_G}
\underline{\mathbf{G}}(\mathbf{r})=\frac{1}{8\pi\eta r}\left[\underline{\mathbf{\hat{I}}}+\mathbf{\hat{r}}\mathbf{\hat{r}}\right], \quad
G(t)=\Theta(t)\mathrm{e}^{-\frac{\mu+\gamma\eta}{\eta}t}. 
\end{equation}
\noindent Here, $\mathbf{\hat{I}}$ denotes the unit matrix, $r=|\mathbf{r}|$, $\mathbf{\hat{r}}=\mathbf{r}/r$, $\mathbf{\hat{r}}\mathbf{\hat{r}}$ is a dyadic product, and $\Theta$ represents the Heaviside step function.

\section{Rotations of the embedded particle}

No-slip conditions are assumed between the surface of the particle and the adjoining viscoelastic environment. We quantify particle rotations by the rotation vector $\bm{\Omega}(t)$. It needs to be interpreted in the same Eulerian way as the memory displacement field $\mathbf{u}(\mathbf{r},t)$ introduced above. That is, $-\bm{\Omega}(t)$ describes the rotation that the particle would perform when the system at time $t$ relaxes back to an unstressed state. If the surrounding medium is not perfectly elastic, also the memory of the previous orientational state will decay over time. In analogy to eq.~(\ref{eq_gamma}) we thus find \cite{puljiz2019memory}
\begin{equation}\label{eq_Omega_W}
\bm{\dot{\Omega}}(t)=\mathbf{W}(t)-\gamma\bm{\Omega}(t).
\end{equation}
\noindent In this expression, $\mathbf{W}(t)$ corresponds to the angular velocity of the particle at time $t$. Thus, the overall physical rotation $\bm{\Omega}^{\mathrm{tot}}(t)$ of the particle starting from the orientational state at a certain time $t_0$ is obtained by time integration of the angular velocity, 
\begin{equation}\label{eq_W}
\bm{\Omega}^{\mathrm{tot}}(t) = \int_{t_0}^{t}\mathbf{W}(t')\mathrm{d}t' +  \bm{\Omega}^{\mathrm{tot}}(t_0) .
\end{equation}

Without loss of generality, we consider the particle to be centered at $\mathbf{r}=\mathbf{0}$. The no-slip surface condition at positions $\mathbf{r}\in\partial V$ on the spherical surface $\partial V$ of the particle reads
\begin{equation}\label{eq_OmegaCrossR}
\bm{\Omega}(t)\times\mathbf{r} = \int_{-\infty}^{\infty}\!\mathrm{d}t'\int_{\partial V}\!\!\mathrm{d}S'\:\underline{\mathbf{G}}(\mathbf{r}-\mathbf{r}',t-t')\cdot\mathbf{f}(\mathbf{r}',t').
\end{equation}
\noindent The left-hand side corresponds to the displacements on the particle surface due to rigid rotations of the particle. They must equal the displacements of the surrounding viscoelastic medium anchored to these positions, as given by the right-hand side of the equation. Here, $\mathbf{f}(\mathbf{r},t)$ represents the surface force density that the particle exerts on the viscoelastic environment at the contact area $\partial V$. In the absence of any other impact on the medium, $\mathbf{f}(\mathbf{r},t)$ is the sole source of distortion. Thus, indeed, the right-hand side of eq.~(\ref{eq_OmegaCrossR}) for $\mathbf{r} \in \partial V$ quantifies the displacements of the viscoelastic medium at the particle surface. Dyadically multiplying this equation by $\mathbf{r}$, integrating over the particle surface $\partial V$, and taking the antisymmetric part of the resulting expression, we obtain
\begin{equation}\label{eq_OmegaT}
\bm{\Omega}(t) = \frac{1}{8\pi\eta a^3}\int_{-\infty}^{\infty}\mathrm{d}t'\:  G(t-t')\mathbf{T}(t').
\end{equation}
\noindent In this expression, $a$ sets the radius of the spherical particle and the torque $\mathbf{T}(t')$ is related to the surface force density $\mathbf{f}(\mathbf{r}',t')$ via
\begin{equation}
\mathbf{T}(t') = \int_{\partial V}\mathrm{d}S'\;\mathbf{r}'\times\mathbf{f}(\mathbf{r}',t'). 
\end{equation}

In our setup, it is the torque $\mathbf{T}(t)$ on the particle that drives the distortions and dynamics of the whole system. This torque is applied onto the particle from outside. It is \textit{not} the torque by which the surrounding medium acts on the particle. The causal chain works in the opposite direction: an external torque is imposed on the particle, the particle transmits this torque to the surrounding viscoelastic medium through the surface anchoring, and as a consequence its environment is set into motion and/or gets distorted. 

We may rescale the relaxation parameter $\gamma$ to obtain the dimensionless number $\mathcal{V}=\gamma\eta/\mu$ \cite{puljiz2019memory}. For a constant imposed torque $\mathbf{T}$, eqs.~(\ref{eq_G}), (\ref{eq_Omega_W}), and (\ref{eq_OmegaT}) imply the particle angular velocity
\begin{equation}
\mathbf{W}(t) = \frac{1}{8\pi\eta a^3}\frac{\mathcal{V}}{1+\mathcal{V}}\,\mathbf{T}. 
\end{equation}
\noindent For reversibly deformable, perfectly elastic, solid-like media, this expression gives the correct limit $\mathbf{W}(t)\rightarrow\mathbf{0}$ for $\mathcal{V}\rightarrow0$. Similarly, in the absence of any elastic contribution, for perfectly viscous fluid media, we correctly recover $\mathbf{W}(t)\rightarrow\mathbf{T}/8\pi\eta a^3$ for $\mathcal{V}\rightarrow\infty$. This expression correctly reproduces Stokes' rotational law for low-Reynolds-number hydrodynamics of incompressible fluids \cite{dhont}. In fact, under these circumstances, the expression in the limit $\mathcal{V}\rightarrow\infty$ also applies for time-dependent torques $\mathbf{T}(t)$.

\section{Magnetically induced particle reorientations}

As outlined above, we consider time-dependent torques that act from outside without any physical contact on individual spherical particles embedded in continuous viscoelastic media. To realize such settings magnetic systems represent natural candidates. Similar situations have been addressed in several experimental, theoretical, and simulation studies in the field of magnetic microrheology \cite{roeder2012shear,roeben2014magnetic,hess2020scale, kreissl2021frequency}. There, information on the dynamics of the viscoelastic medium is extracted from the configurational response of embedded magnetic particles to time-dependent external magnetic fields. Naturally, in many experimental situations the response of more than a single particle is monitored simultaneously. Nevertheless, for low particle concentrations, mutual particle interactions are neglected and the single-particle response still provides a reasonable measure. 

We concentrate on a uniformly magnetized, homogeneous, magnetically hard particle in a nonmagnetic viscoelastic environment. That is, the spherical particle features a permanent magnetic dipole moment $\mathbf{m}(t)$ of constant magnitude $m=\|\mathbf{m}(t)\|$. Therefore, $\mathbf{m}(t)=m\mathbf{\hat{m}}(t)$, where $\mathbf{\hat{m}}(t)$ is rigidly and permanently anchored to the particle frame. Consequently, changes in magnetic properties only result from particle reorientations \cite{PhysRevE.97.032610,kreissl2021frequency}. 

The torque $\mathbf{T}(t)$ imposed on the magnetic dipole moment $\mathbf{m}(t)$ in a spatially homogeneous external magnetic field $\mathbf{B}(t)$ and thus on the whole magnetic particle is given by
\begin{equation}\label{eq_TmB}
\mathbf{T}(t)=\mathbf{m}(t)\times\mathbf{B}(t). 
\end{equation}
\noindent Since $m$ is constant and the magnetic moment is rigidly anchored to the particle frame, the only changes in $\mathbf{m}(t)$ result from rotations of the whole particle by the angular velocity $\mathbf{W}(t)$,
\begin{equation}\label{eq_m}
\frac{\mathrm{d}\mathbf{m}(t)}{\mathrm{d}t} = \mathbf{W}(t)\times\mathbf{m}(t). 
\end{equation}
\noindent Inserting eq.~(\ref{eq_TmB}) into eq.~(\ref{eq_OmegaT}) and the latter into eq.~(\ref{eq_Omega_W}), we obtain an expression for the particle angular velocity $\mathbf{W}(t)$, which via eq.~(\ref{eq_m}) determines the dynamics of the magnetic moment $\mathbf{m}(t)$. Along these lines, we find
\begin{eqnarray}
\frac{\mathrm{d}\mathbf{m}}{\mathrm{d}t} &=&
\frac{1}{8\pi\eta a^3}\bigg\{
\left[\mathbf{m}(t)\times\mathbf{B}(t)\right]\times\mathbf{m}(t)
\nonumber\\
&&
-\frac{\mu}{\eta}\mathrm{e}^{-\frac{\mu+\gamma\eta}{\eta}t}
\!\!
\int_{-\infty}^t \!\!\!\mathrm{d}t'\,
\mathrm{e}^{\frac{\mu+\gamma\eta}{\eta}t'}\!
\left[\mathbf{m}(t')\hspace{-1pt}\times\hspace{-1pt}\mathbf{B}(t')\right]\hspace{-1pt}\times\hspace{-1pt}\mathbf{m}(t)
\bigg\}.
\nonumber\\&&
\end{eqnarray}

Next, we parameterize
\begin{equation}\label{eq_param}
\mathbf{\hat{m}}(t)=\left(\begin{array}{c}
\cos\varphi(t) \\ 0 \\ \sin\varphi(t) \end{array}\right), \qquad
\mathbf{B}(t) = B(t)\left(\begin{array}{c}
0 \\ 0 \\ 1 \end{array}\right).
\end{equation}
\noindent This leads us to the general dynamic response of the particle, quantified by the time evolution equation for the orientational angle $\varphi(t)$,
\begin{eqnarray}
\frac{\mathrm{d}\varphi(t)}{\mathrm{d}t} &=&
\frac{m}{8\pi\eta a^3}\bigg\{ B(t)\cos\varphi(t)
\nonumber\\
&&
{}-\frac{\mu}{\eta}\mathrm{e}^{-\frac{\mu+\gamma\eta}{\eta}t}
\int_{-\infty}^t\mathrm{d}t'\:
\mathrm{e}^{\frac{\mu+\gamma\eta}{\eta}t'}\cos\varphi(t')B(t')\bigg\}.
\nonumber\\
&& 
\end{eqnarray}
\noindent Additionally, we specify the external magnetic field to oscillate with a frequency $\omega$,
\begin{equation}\label{eq_Bt}
B(t)=B\cos(\omega t). 
\end{equation}
\noindent Next, we rescale the variables to dimensionless ones, setting $t=\tilde{t}\eta/\mu$, $\omega=\tilde{\omega}\mu/\eta$, and $\varphi(t)=\tilde{\varphi}(\tilde{t})$. Thus the dynamic equation for the orientation angle becomes
\begin{eqnarray}
\frac{\mathrm{d}\tilde{\varphi}}{\mathrm{d}\tilde{t}} &=&
\frac{mB}{8\pi\mu a^3}\bigg\{\cos(\tilde{\omega}\tilde{t})\cos\tilde{\varphi}(\tilde{t})
\nonumber\\
&&
{}-\mathrm{e}^{-(1+\mathcal{V})\tilde{t}}
\int_{-\infty}^{\tilde{t}}\mathrm{d}\tilde{t}'\:
\mathrm{e}^{(1+\mathcal{V})\tilde{t}'}\cos\tilde{\varphi}(\tilde{t}')
\cos(\tilde{\omega}\tilde{t}'). 
\nonumber\\&&
\end{eqnarray}

From here, we assume $|B|$ to be small enough so that only the resulting linear response \cite{kubo1991statistical,zwanzig2001nonequilibrium} of the system to the external magnetic field needs to be addressed. For this purpose, we set
\begin{equation}\label{eq_defdelphi}
\tilde{\varphi}(\tilde{t})=\varphi_0+\delta\tilde{\varphi}(\tilde{t}),
\end{equation}
\noindent where $\varphi_0$ quantifies the equilibrium orientation of $\mathbf{\hat{m}}$ in the absence of any external magnetic field. This leads us to
\begin{eqnarray}
\delta\tilde{\varphi}(\tilde{t}) &=&
\frac{mB\cos\varphi_0}{8\pi\mu a^3}\bigg\{
\frac{1}{\tilde{\omega}^2+(1+\mathcal{V})^2}\cos(\tilde{\omega}\tilde{t})
\nonumber\\
&&{}
\qquad
+\frac{1}{\tilde{\omega}}
\left[1-\frac{1+\mathcal{V}}{\tilde{\omega}^2+(1+\mathcal{V})^2}\right]
\sin(\tilde{\omega}\tilde{t})\bigg\}.
\qquad
\label{eq_delphitilde}
\end{eqnarray}

\section{Distortions and flows of the viscoelastic environment}

Addressing time-dependent torques $\mathbf{T}(t)$ acting on the rigid spherical particle centered around $\mathbf{r}=\mathbf{0}$, we now calculate the elastic distortions and flows induced through the no-slip surface coupling in the surrounding viscoelastic medium. Under the given circumstances and assumptions, the resulting time-dependent memory displacement field in the viscoelastic environment for $r\geq a$ reads \cite{puljiz2019memory}
\begin{equation}
\mathbf{u}(\mathbf{r},t) = {}-\frac{1}{2}\int_{-\infty}^{\infty}\mathrm{d}t'\:G(t-t')\,\mathbf{T}(t')\cdot\left[\bm{\nabla}\times
\underline{\mathbf{G}}(\mathbf{r})\right]. \;
\end{equation}
\noindent In the static limit ($t\rightarrow\infty$) of a constant torque $\mathbf{T}$ in a perfectly elastic solid-like environment, this expression turns into the correct form $\mathbf{u}(\mathbf{r})={}-(\mathbf{T}\times\bm{\nabla})\cdot\underline{\mathbf{G}}(\mathbf{r})\eta/2\mu$ \cite{phan1993rigid,phan1994load, puljiz2017forces,puljiz2019displacement}. 

Again involving the magnetic torque given by eq.~(\ref{eq_TmB}) together with the parameterizations listed in eq.~(\ref{eq_param}), we find
\begin{eqnarray}
\mathbf{u}(\mathbf{r},t) &=&
\frac{1}{2}\,m\,\mathbf{\hat{y}}\cdot\left[\bm{\nabla}\times
\underline{\mathbf{G}}(\mathbf{r})\right]
\mathrm{e}^{-\frac{\mu+\gamma\eta}{\eta}t}
\nonumber\\
&&
\qquad{}\times
\int_{-\infty}^t\mathrm{d}t'\:
\mathrm{e}^{\frac{\mu+\gamma\eta}{\eta}t'}B(t')\cos\varphi(t').
\qquad
\end{eqnarray}
\noindent Here, the exposed role of the unit vector $\mathbf{\hat{y}}$ results from the choice of our coordinate system as implied by eq.~(\ref{eq_param}). There, we chose the $x$-$z$-plane to coincide with the rotational plane of $\mathbf{\hat{m}}(t)$, while $\mathbf{B}(t)\parallel\mathbf{\hat{z}}$. The simultaneously induced flow $\mathbf{v}(\mathbf{r},t)$ in the viscoelastic environment follows via eq.~(\ref{eq_gamma}) as $\mathbf{v}(\mathbf{r},t)=\mathbf{\dot{u}}(\mathbf{r},t)+\gamma\mathbf{u}(\mathbf{r},t)$.

Next, we involve the time dependence of the magnetic field as given by eq.~(\ref{eq_Bt}). Switching again to the rescaled variables introduced above and confining ourselves to the linear response as indicated by eq.~(\ref{eq_defdelphi}), we obtain
\begin{equation}\label{eq_utilde}
\tilde{\mathbf{u}}(\mathbf{r},\tilde{t})=
\frac{Bm\cos\varphi_0}{8\pi\mu}
\,
\frac{(1+\mathcal{V})\cos(\tilde{\omega}\tilde{t})
+\tilde{\omega}\sin(\tilde{\omega}\tilde{t})}
{\tilde{\omega}^2+(1+\mathcal{V})^2}
\,
\frac{1}{r^3}\mathbf{r}\times\mathbf{\hat{y}}.
\end{equation}
\noindent In this expression, we defined $\tilde{\mathbf{u}}(\mathbf{r},\tilde{t})=\mathbf{u}(\mathbf{r},t)$. An additional rescaling of lengths by $(Bm/8\pi\mu)^{(1/3)}$ would be possible, but we wish to keep $B$ and $m$ explicit here. 

We define the rescaled flow fields induced in the viscoelastic medium by the external torque on the particle as $\tilde{\mathbf{v}}(\mathbf{r},\tilde{t})=\mathbf{v}(\mathbf{r},t)\eta/\mu$. Then, via eq.~(\ref{eq_gamma}), we find $\tilde{\mathbf{v}}(\mathbf{r},\tilde{t})=\tilde{\mathbf{\dot{\mathbf{u}}}}(\mathbf{r},\tilde{t})+\mathcal{V}\tilde{\mathbf{u}}(\mathbf{r},\tilde{t})$. Inserting eq.~(\ref{eq_utilde}) leads to
\begin{align}
\tilde{\mathbf{v}}(\mathbf{r},\tilde{t}) =& \frac{Bm\cos\varphi_0}{8\pi\mu}
\nonumber\\
&
\times\frac{\left[\tilde{\omega}^2+\mathcal{V}(1+\mathcal{V})\right]\cos(\tilde{\omega}\tilde{t})-\tilde{\omega}\sin(\tilde{\omega}\tilde{t})}
{\tilde{\omega}^2+(1+\mathcal{V})^2}
\frac{1}{r^3}\mathbf{r}\times\mathbf{\hat{y}}.
\nonumber\\&
\label{eq_vtilde}
\end{align}

Thus, in the limits of a perfectly elastic surrounding medium ($\mathcal{V}=0$) and of quasistatic dynamics ($\tilde{\omega}\rightarrow0$), we find an elastic displacement field in phase with the external magnetic driving field, $\tilde{\mathbf{u}}(\mathbf{r},\tilde{t})\sim\cos(\tilde{\omega}\tilde{t})\,\mathbf{r}\times\mathbf{\hat{y}}/r^3$. Conversely, in the limit of a perfectly viscous fluid ($\mathcal{V}\rightarrow\infty$), the displacement memory field vanishes, $\tilde{\mathbf{u}}(\mathbf{r},\tilde{t})\rightarrow0$, and the induced flow is in phase with the magnetic stimulus, $\tilde{\mathbf{v}}(\mathbf{r},\tilde{t})\sim\cos(\tilde{\omega}\tilde{t})\,\mathbf{r}\times\mathbf{\hat{y}}/r^3$.

\section{Linear response function and magnetic susceptibility}

Rewriting the external magnetic stimulus in eq.~(\ref{eq_Bt}) as
\begin{equation}
B(\tilde{t})=\mathrm{Re}\left\{B\,\mathrm{e}^{\mathrm{i}\tilde{\omega}\tilde{t}}\right\}, 
\end{equation}
\noindent where $\mathrm{Re}$ indicates the real part, we may define the linear response function $\chi_{\tilde{\varphi}}(\tilde{\omega})=\chi_{\tilde{\varphi}}'(\tilde{\omega})-\mathrm{i}\chi_{\tilde{\varphi}}''(\tilde{\omega})$ of the angular response $\delta\tilde{\varphi}(\tilde{t})$ to the stimulus $B(\tilde{t})$ via
\begin{equation}
\delta\tilde{\varphi}(\tilde{t})\,=\,\mathrm{Re}\left\{\chi_{\tilde{\varphi}}(\tilde{\omega})\,B\,\mathrm{e}^{\mathrm{i}\tilde{\omega}\tilde{t}}\right\}. 
\end{equation}
\noindent From eq.~(\ref{eq_delphitilde}), we find
\begin{eqnarray}
\chi_{\tilde{\varphi}}'(\tilde{\omega}) &=&
\frac{m\cos\varphi_0}{8\pi\mu a^3}\,\frac{1}{\tilde{\omega}^2+(1+\mathcal{V})^2},
\label{eq_chiphire}
\\
\chi_{\tilde{\varphi}}''(\tilde{\omega}) &=&
\frac{m\cos\varphi_0}{8\pi\mu a^3}\,\frac{1}{\tilde{\omega}}
\left(1-\frac{1+\mathcal{V}}{\tilde{\omega}^2+(1+\mathcal{V})^2}\right).
\label{eq_chiphiim}
\end{eqnarray}

In experiments, larger systems may be studied in sufficiently macroscopic measurements. In such settings one would probably investigate samples of not just one enclosed magnetic particle but rather many of them. The quantity then to be determined is typically the induced magnetization $\mathbf{M}(t)$ along the external magnetic field direction. In our case, since $\mathbf{B}(t)\parallel\mathbf{\hat{z}}$, this reduces our considerations to $M_z(t)$. The linear response function usually referred to in this context is the magnetic susceptibility $\chi_{\tilde{\omega}}$. By convention, it relates the response $\mathbf{M}(t)$ to the stimulus magnetic field $\mathbf{H}(t)$. Thus, an additional factor of $\mu_0$, denoting the magnetic vacuum permeability, is included. 

To address this situation, we need to know how the magnetic particles are initially organized in the system. Their density is assumed to be low enough so that mutual interactions between the particles can be neglected. We denote the particle number density as $c$. One basic scenario would imply all particles to initially feature a magnetization perpendicular to the $\mathbf{\hat{z}}$-direction. A corresponding situation could be achieved if the magnetic particles allow for a ``(re)programming'' of the directions of their magnetic moments by strong external magnetic fields. Then, first, a strong magnetic field along, \textit{e.g.}, the $\mathbf{\hat{x}}$-direction could be applied, before the magnetic response along the $\mathbf{\hat{z}}$-direction is probed using significantly weaker fields. In that case, $\varphi_0\approx0$ for all particles. This implies $M_z(\tilde{t})\approx c\,m\,\delta\tilde{\varphi}(\tilde{t})$, while simultaneously
\begin{equation}
M_z(\tilde{t})=\mathrm{Re}\left\{ \chi(\tilde{\omega})\frac{1}{\mu_0}B
\,\mathrm{e}^{\mathrm{i}\tilde{\omega}\tilde{t}} \right\}. 
\end{equation}
\noindent Thus, we find $\chi(\tilde{\omega})=\mu_0cm\,\chi_{\tilde{\varphi}}(\tilde{\omega})$, where $\chi_{\tilde{\varphi}}(\tilde{\omega})$ can be read off from eqs.~(\ref{eq_chiphire}) and (\ref{eq_chiphiim}), setting $\cos(\varphi_0)=1$. 

If instead we focus on a situation of a rather random initial distribution of the orientations of the magnetic moments, we need to average the above expressions over the corresponding angular distribution of $\varphi_0$. Since $\langle\cos^2\varphi_0\rangle=2/3$, we need to multiply the above expressions by this factor. We rescale $\chi({\tilde{\omega}})=\tilde{\chi}(\tilde{\omega})\,\mu_0cm^2/12\pi\mu a^3$ to obtain
\begin{eqnarray}
\tilde{\chi}'(\tilde{\omega}) &=&
\frac{1}{\tilde{\omega}^2+(1+\mathcal{V})^2},
\label{eq_chire}
\\ 
\tilde{\chi}''(\tilde{\omega}) &=&
\frac{1}{\tilde{\omega}}\left(1-
\frac{1+\mathcal{V}}{\tilde{\omega}^2+(1+\mathcal{V})^2}\right).
\label{eq_chiim}
\end{eqnarray}
\noindent For a perfectly elastic embedding medium ($\mathcal{V}=0$), we find in the limit of quasistatic dynamics ($\tilde{\omega}\rightarrow0$) an in-phase magnetic response, as $\tilde{\chi}'(\tilde{\omega})\rightarrow1$ and $\tilde{\chi}''(\tilde{\omega})\rightarrow0$. Simultaneously, for a perfectly viscous surrounding fluid ($\mathcal{V}\rightarrow\infty$), we obtain $\tilde{\chi}'(\tilde{\omega})\rightarrow0$ while $\tilde{\chi}''(\tilde{\omega})\rightarrow1/\tilde{\omega}$. 

\begin{figure}
\includegraphics[width=\columnwidth]{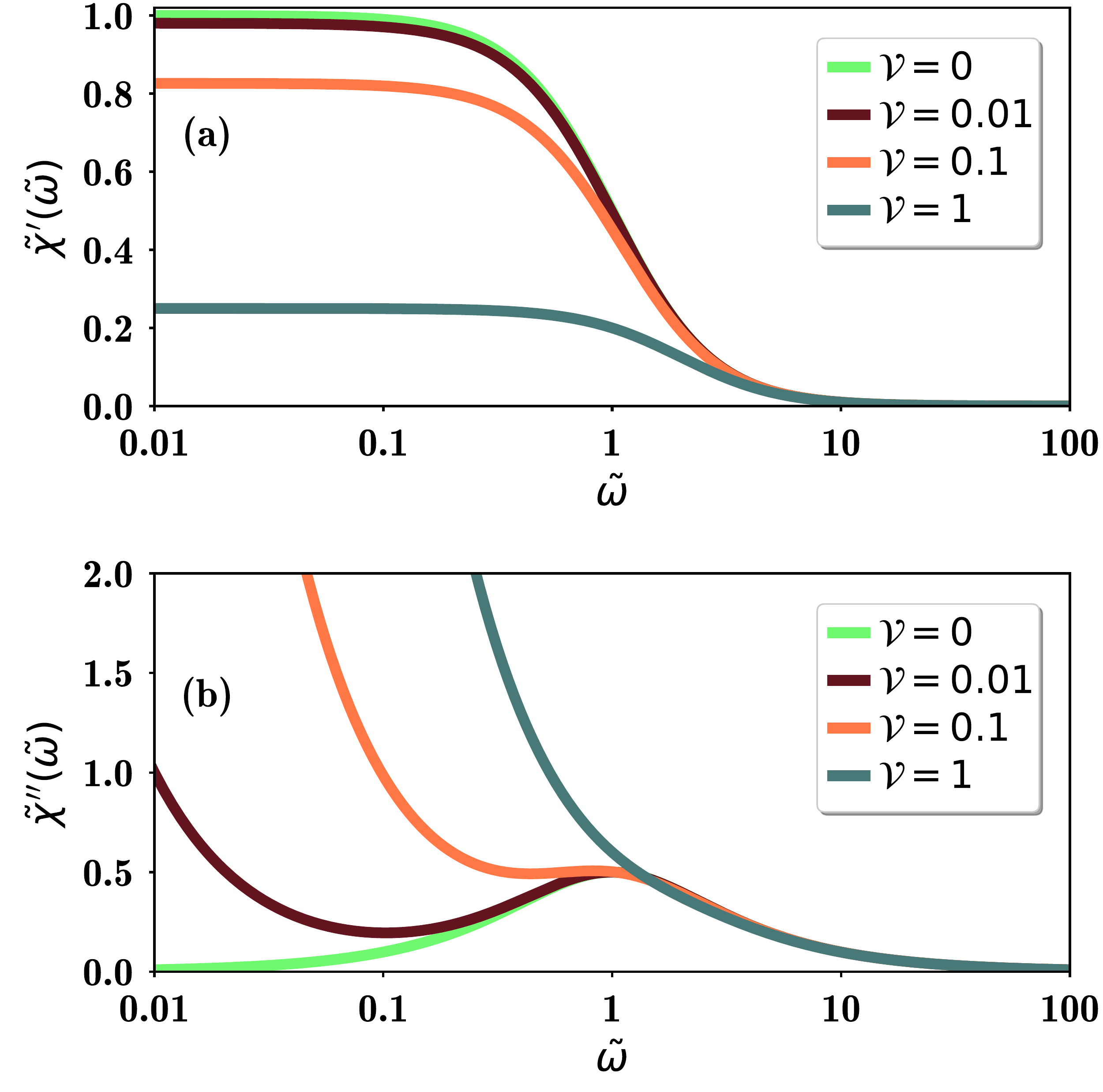}
\caption{Rescaled magnetic susceptibility $\tilde{\chi}(\tilde{\omega})=\tilde{\chi}'(\tilde{\omega})-\mathrm{i}\tilde{\chi}''(\tilde{\omega})$ as a function of the scaled frequency $\tilde{\omega}$, see eqs.~(\ref{eq_chire}) and (\ref{eq_chiim}), for different values of the rescaled relaxation parameter $\mathcal{V}$. While $\mathcal{V}=0$ identifies reversible elastic deformations, $\mathcal{V}\rightarrow\infty$ corresponds to a viscous fluid. When varying $\mathcal{V}$, the shape of the curves remains qualitatively similar for $\tilde{\chi}'(\tilde{\omega})$ in (a). Conversely, the terminal flow behavior for $\mathcal{V}>0$ is reflected by the divergence of $\tilde{\chi}''(\tilde{\omega})$ for $\tilde{\omega}\rightarrow0$ in (b), while $\tilde{\chi}''(\tilde{\omega}\rightarrow 0)\rightarrow 0$ for $\mathcal{V}=0$, \textit{i.e.}, for genuinely elastic systems.}
\label{fig_chi}
\end{figure}

Further details are displayed in fig.~\ref{fig_chi}. 
We note that $\tilde{\chi}'(\tilde{\omega})$ and $\tilde{\chi}''(\tilde{\omega})$ show very different trends depending on the nature of the embedding medium. First, concerning $\tilde{\chi}'(\tilde{\omega})$, genuinely elastic media ($\mathcal{V}=0$) and those featuring terminal flow ($\mathcal{V}>0$) lead to qualitatively similar response. This becomes even more evident upon further rescaling in eq.~(\ref{eq_chire}) $\tilde{\omega}$ by $1+\mathcal{V}$ and $\tilde{\chi}'(\tilde{\omega})$ by $1/(1+\mathcal{V})^2$. Then, all curves collapse. Conversely, the properties of $\tilde{\chi}''(\tilde{\omega})$ become qualitatively different in these two distinct situations, particularly concerning the low-frequency limit. While $\tilde{\chi}''(\tilde{\omega}\rightarrow0)=0$ for elastic media ($\mathcal{V}=0$), we find $\tilde{\chi}''(\tilde{\omega}\rightarrow0)\rightarrow\infty$ for fluid-like behavior ($\mathcal{V}>0$). These two situations correctly reflect the low-frequency limits of genuinely elastic media and truly fluid systems, respectively \cite{strobl2007physics}. In the fluid case at low enough frequencies, a persistent rotation of the particle combined with a terminal flow of the surrounding medium becomes possible. Still, we should remember that our evaluation is restricted to the regime of linear response and angular deviations $\delta\tilde{\varphi}(\tilde{t})$  must remain of sufficiently small magnitude. Thus, the amplitude of the magnetic field must be confined accordingly.

Importantly, in their unscaled versions, eqs.~(\ref{eq_chire}) and (\ref{eq_chiim}) allow to draw conclusions on the actual values of $\mu$, $\eta$, and $\gamma$ when they are fitted to actual experimental results. This enables statements on these parameters in a type of microrheological approach. During practical applications, one should recall in this context the restriction of our considerations to overdamped dynamics on the considered micro- to mesoscopic scales. 

Finally, if the orientations of the magnetic moments are initially randomly distributed, so are their projections into the plane perpendicular to $\mathbf{B}(t)$. Thus, the overall averaged memory displacement field $\tilde{\mathbf{u}}(\mathbf{r},\tilde{t})$ and flow field $\tilde{\mathbf{v}}(\mathbf{r},\tilde{t})$ induced by all particles together vanish. This follows from eqs.~(\ref{eq_utilde}) and (\ref{eq_vtilde}), respectively. There, instead of $\mathbf{\hat{y}}$, a randomized orientational unit vector in the $x$-$y$-plane appears, which does not endure orientational averaging.

\section{Conclusions}

Summarizing, we analyzed the response of a rigid spherical particle embedded under no-slip surface conditions in a continuous viscoelastic environment to externally imposed torques. The embedding viscoelastic medium can either feature terminal flow behavior or perfectly elastic reversible deformations. Both situations are addressed by the same formalism, where one relaxational parameter allows to set the nature of the viscoelastic environment. We confined ourselves to incompressible systems and overdamped dynamics, which should apply, for example, to many polymeric synthetic gel-like systems as well as to biological environments. The size of the embedded particles was assumed to be large enough so that thermal fluctuations may be disregarded. 

In view of possible magnetic microrheological applications, we addressed the linear magnetic response to oscillating external magnetic fields for magnetically hard, \textit{i.e.}, magnetically blocked particles serving as particle probes. That is, a permanent magnetic moment is anchored to the particle frame, and magnetic reorientations imply particle rotations. The magnetic frequency-dependent susceptibility function was derived in this case. Accordingly, we provide a theoretical framework that allows to describe the properties of viscoelastic materials by fitting the expressions of the magnetic susceptibility to corresponding magnetic microrheological measurement results. The material response can then be interpreted and described using our theoretical framework, including the calculation of the distortions and flows induced in the environment of the probe particles. 

Beyond the considered framework, our investigations should as well be interesting for the study of self-propelled microswimmers in viscoelastic media \cite{puljiz2019memory, narinder2019active, qi2020enhanced,Zottl.2019}. This especially applies when self-propulsion is induced for magnetic particles by rotating external magnetic fields \cite{martinez2018emergent}. As explained, our work is motivated by a microrheological perspective. Nevertheless, our description within the listed boundaries is equally suitable to characterize macroscopic settings of larger sphere embedded in a viscoelastic environment.

\acknowledgments
The authors acknowledge helpful and stimulating discussions with Patrick Kreissl. 
A.M.M.\ thanks the Deut\-sche For\-schungs\-ge\-mein\-schaft (German Research Foundation, DFG) for support through the Heisenberg Grant ME 3571/4-1.

\bibliography{References}

\begin{thebibliography}{40}
\expandafter\ifx\csname natexlab\endcsname\relax\def\natexlab#1{#1}\fi
\expandafter\ifx\csname bibnamefont\endcsname\relax
  \def\bibnamefont#1{#1}\fi
\expandafter\ifx\csname bibfnamefont\endcsname\relax
  \def\bibfnamefont#1{#1}\fi
\expandafter\ifx\csname citenamefont\endcsname\relax
  \def\citenamefont#1{#1}\fi
\expandafter\ifx\csname url\endcsname\relax
  \def\url#1{\texttt{#1}}\fi
\expandafter\ifx\csname urlprefix\endcsname\relax\def\urlprefix{URL }\fi
\providecommand{\bibinfo}[2]{#2}
\providecommand{\eprint}[2][]{\url{#2}}

\bibitem[{\citenamefont{Huang et~al.}(2016)\citenamefont{Huang, Pessot, Cremer,
  Weeber, Holm, Nowak, Odenbach, Menzel, and Auernhammer}}]{huang2016buckling}
\bibinfo{author}{\bibfnamefont{S.}~\bibnamefont{Huang}},
  \bibinfo{author}{\bibfnamefont{G.}~\bibnamefont{Pessot}},
  \bibinfo{author}{\bibfnamefont{P.}~\bibnamefont{Cremer}},
  \bibinfo{author}{\bibfnamefont{R.}~\bibnamefont{Weeber}},
  \bibinfo{author}{\bibfnamefont{C.}~\bibnamefont{Holm}},
  \bibinfo{author}{\bibfnamefont{J.}~\bibnamefont{Nowak}},
  \bibinfo{author}{\bibfnamefont{S.}~\bibnamefont{Odenbach}},
  \bibinfo{author}{\bibfnamefont{A.~M.} \bibnamefont{Menzel}},
  \bibnamefont{and} \bibinfo{author}{\bibfnamefont{G.~K.}
  \bibnamefont{Auernhammer}}, \bibinfo{journal}{Soft Matter}
  \textbf{\bibinfo{volume}{12}}, \bibinfo{pages}{228} (\bibinfo{year}{2016}).

\bibitem[{\citenamefont{Puljiz et~al.}(2016)\citenamefont{Puljiz, Huang,
  Auernhammer, and Menzel}}]{puljiz2016forces}
\bibinfo{author}{\bibfnamefont{M.}~\bibnamefont{Puljiz}},
  \bibinfo{author}{\bibfnamefont{S.}~\bibnamefont{Huang}},
  \bibinfo{author}{\bibfnamefont{G.~K.} \bibnamefont{Auernhammer}},
  \bibnamefont{and} \bibinfo{author}{\bibfnamefont{A.~M.}
  \bibnamefont{Menzel}}, \bibinfo{journal}{Phys. Rev. Lett.}
  \textbf{\bibinfo{volume}{117}}, \bibinfo{pages}{238003}
  (\bibinfo{year}{2016}).

\bibitem[{\citenamefont{MacKintosh and
  Schmidt}(1999)}]{mackintosh1999microrheology}
\bibinfo{author}{\bibfnamefont{F.~C.} \bibnamefont{MacKintosh}}
  \bibnamefont{and} \bibinfo{author}{\bibfnamefont{C.~F.}
  \bibnamefont{Schmidt}}, \bibinfo{journal}{Curr. Opin. Colloid Interface Sci.}
  \textbf{\bibinfo{volume}{4}}, \bibinfo{pages}{300} (\bibinfo{year}{1999}).

\bibitem[{\citenamefont{Crocker et~al.}(2000)\citenamefont{Crocker, Valentine,
  Weeks, Gisler, Kaplan, Yodh, and Weitz}}]{crocker2000two}
\bibinfo{author}{\bibfnamefont{J.~C.} \bibnamefont{Crocker}},
  \bibinfo{author}{\bibfnamefont{M.~T.} \bibnamefont{Valentine}},
  \bibinfo{author}{\bibfnamefont{E.~R.} \bibnamefont{Weeks}},
  \bibinfo{author}{\bibfnamefont{T.}~\bibnamefont{Gisler}},
  \bibinfo{author}{\bibfnamefont{P.~D.} \bibnamefont{Kaplan}},
  \bibinfo{author}{\bibfnamefont{A.~G.} \bibnamefont{Yodh}}, \bibnamefont{and}
  \bibinfo{author}{\bibfnamefont{D.~A.} \bibnamefont{Weitz}},
  \bibinfo{journal}{Phys. Rev. Lett.} \textbf{\bibinfo{volume}{85}},
  \bibinfo{pages}{888} (\bibinfo{year}{2000}).

\bibitem[{\citenamefont{Levine and Lubensky}(2000)}]{levine2000one}
\bibinfo{author}{\bibfnamefont{A.~J.} \bibnamefont{Levine}} \bibnamefont{and}
  \bibinfo{author}{\bibfnamefont{T.~C.} \bibnamefont{Lubensky}},
  \bibinfo{journal}{Phys. Rev. Lett.} \textbf{\bibinfo{volume}{85}},
  \bibinfo{pages}{1774} (\bibinfo{year}{2000}).

\bibitem[{\citenamefont{Waigh}(2005)}]{waigh2005microrheology}
\bibinfo{author}{\bibfnamefont{T.~A.} \bibnamefont{Waigh}},
  \bibinfo{journal}{Rep. Prog. Phys.} \textbf{\bibinfo{volume}{68}},
  \bibinfo{pages}{685} (\bibinfo{year}{2005}).

\bibitem[{\citenamefont{Wilhelm}(2008)}]{wilhelm2008out}
\bibinfo{author}{\bibfnamefont{C.}~\bibnamefont{Wilhelm}},
  \bibinfo{journal}{Phys. Rev. Lett.} \textbf{\bibinfo{volume}{101}},
  \bibinfo{pages}{028101} (\bibinfo{year}{2008}).

\bibitem[{\citenamefont{Wirtz}(2009)}]{wirtz2009particle}
\bibinfo{author}{\bibfnamefont{D.}~\bibnamefont{Wirtz}},
  \bibinfo{journal}{Annu. Rev. Biophys.} \textbf{\bibinfo{volume}{38}},
  \bibinfo{pages}{301} (\bibinfo{year}{2009}).

\bibitem[{\citenamefont{Squires and Mason}(2010)}]{squires2010fluid}
\bibinfo{author}{\bibfnamefont{T.~M.} \bibnamefont{Squires}} \bibnamefont{and}
  \bibinfo{author}{\bibfnamefont{T.~G.} \bibnamefont{Mason}},
  \bibinfo{journal}{Annu. Rev. Fluid Mech.} \textbf{\bibinfo{volume}{42}},
  \bibinfo{pages}{413} (\bibinfo{year}{2010}).

\bibitem[{\citenamefont{Puertas and
  Voigtmann}(2014)}]{puertas2014microrheology}
\bibinfo{author}{\bibfnamefont{A.~M.} \bibnamefont{Puertas}} \bibnamefont{and}
  \bibinfo{author}{\bibfnamefont{T.}~\bibnamefont{Voigtmann}},
  \bibinfo{journal}{J. Phys.: Condens. Matter} \textbf{\bibinfo{volume}{26}},
  \bibinfo{pages}{243101} (\bibinfo{year}{2014}).

\bibitem[{\citenamefont{Paul et~al.}(2019)\citenamefont{Paul, Kundu, and
  Banerjee}}]{paul2019active}
\bibinfo{author}{\bibfnamefont{S.}~\bibnamefont{Paul}},
  \bibinfo{author}{\bibfnamefont{A.}~\bibnamefont{Kundu}}, \bibnamefont{and}
  \bibinfo{author}{\bibfnamefont{A.}~\bibnamefont{Banerjee}},
  \bibinfo{journal}{J. Phys. Commun.} \textbf{\bibinfo{volume}{3}},
  \bibinfo{pages}{035002} (\bibinfo{year}{2019}).

\bibitem[{\citenamefont{Bausch et~al.}(1999)\citenamefont{Bausch, M{\"o}ller,
  and Sackmann}}]{bausch1999measurement}
\bibinfo{author}{\bibfnamefont{A.~R.} \bibnamefont{Bausch}},
  \bibinfo{author}{\bibfnamefont{W.}~\bibnamefont{M{\"o}ller}},
  \bibnamefont{and} \bibinfo{author}{\bibfnamefont{E.}~\bibnamefont{Sackmann}},
  \bibinfo{journal}{Biophys. J.} \textbf{\bibinfo{volume}{76}},
  \bibinfo{pages}{573} (\bibinfo{year}{1999}).

\bibitem[{\citenamefont{Raikher and Rusakov}(2006)}]{raikher2006magnetic}
\bibinfo{author}{\bibfnamefont{Y.~L.} \bibnamefont{Raikher}} \bibnamefont{and}
  \bibinfo{author}{\bibfnamefont{V.~V.} \bibnamefont{Rusakov}},
  \bibinfo{journal}{J. Magn. Magn. Mater.} \textbf{\bibinfo{volume}{300}},
  \bibinfo{pages}{e229} (\bibinfo{year}{2006}).

\bibitem[{\citenamefont{Roeder et~al.}(2012)\citenamefont{Roeder, Bender,
  Tsch{\"o}pe, Birringer, and Schmidt}}]{roeder2012shear}
\bibinfo{author}{\bibfnamefont{L.}~\bibnamefont{Roeder}},
  \bibinfo{author}{\bibfnamefont{P.}~\bibnamefont{Bender}},
  \bibinfo{author}{\bibfnamefont{A.}~\bibnamefont{Tsch{\"o}pe}},
  \bibinfo{author}{\bibfnamefont{R.}~\bibnamefont{Birringer}},
  \bibnamefont{and} \bibinfo{author}{\bibfnamefont{A.~M.}
  \bibnamefont{Schmidt}}, \bibinfo{journal}{J. Polym. Sci. B}
  \textbf{\bibinfo{volume}{50}}, \bibinfo{pages}{1772} (\bibinfo{year}{2012}).

\bibitem[{\citenamefont{Roeben et~al.}(2014)\citenamefont{Roeben, Roeder,
  Teusch, Effertz, Deiters, and Schmidt}}]{roeben2014magnetic}
\bibinfo{author}{\bibfnamefont{E.}~\bibnamefont{Roeben}},
  \bibinfo{author}{\bibfnamefont{L.}~\bibnamefont{Roeder}},
  \bibinfo{author}{\bibfnamefont{S.}~\bibnamefont{Teusch}},
  \bibinfo{author}{\bibfnamefont{M.}~\bibnamefont{Effertz}},
  \bibinfo{author}{\bibfnamefont{U.~K.} \bibnamefont{Deiters}},
  \bibnamefont{and} \bibinfo{author}{\bibfnamefont{A.~M.}
  \bibnamefont{Schmidt}}, \bibinfo{journal}{Colloid Polym. Sci.}
  \textbf{\bibinfo{volume}{292}}, \bibinfo{pages}{2013} (\bibinfo{year}{2014}).

\bibitem[{\citenamefont{Hess et~al.}(2020)\citenamefont{Hess, Gratz, Remmer,
  Webers, Landers, Borin, Ludwig, Wende, Odenbach, Tsch{\"o}pe
  et~al.}}]{hess2020scale}
\bibinfo{author}{\bibfnamefont{M.}~\bibnamefont{Hess}},
  \bibinfo{author}{\bibfnamefont{M.}~\bibnamefont{Gratz}},
  \bibinfo{author}{\bibfnamefont{H.}~\bibnamefont{Remmer}},
  \bibinfo{author}{\bibfnamefont{S.}~\bibnamefont{Webers}},
  \bibinfo{author}{\bibfnamefont{J.}~\bibnamefont{Landers}},
  \bibinfo{author}{\bibfnamefont{D.}~\bibnamefont{Borin}},
  \bibinfo{author}{\bibfnamefont{F.}~\bibnamefont{Ludwig}},
  \bibinfo{author}{\bibfnamefont{H.}~\bibnamefont{Wende}},
  \bibinfo{author}{\bibfnamefont{S.}~\bibnamefont{Odenbach}},
  \bibinfo{author}{\bibfnamefont{A.}~\bibnamefont{Tsch{\"o}pe}},
  \bibnamefont{et~al.}, \bibinfo{journal}{Soft Matter}
  \textbf{\bibinfo{volume}{16}}, \bibinfo{pages}{7562} (\bibinfo{year}{2020}).

\bibitem[{\citenamefont{Filipcsei et~al.}(2007)\citenamefont{Filipcsei,
  Csetneki, Szil{\'a}gyi, and Zr{\'\i}nyi}}]{filipcsei2007magnetic}
\bibinfo{author}{\bibfnamefont{G.}~\bibnamefont{Filipcsei}},
  \bibinfo{author}{\bibfnamefont{I.}~\bibnamefont{Csetneki}},
  \bibinfo{author}{\bibfnamefont{A.}~\bibnamefont{Szil{\'a}gyi}},
  \bibnamefont{and}
  \bibinfo{author}{\bibfnamefont{M.}~\bibnamefont{Zr{\'\i}nyi}},
  \bibinfo{journal}{Adv. Polym. Sci.} \textbf{\bibinfo{volume}{107}},
  \bibinfo{pages}{137} (\bibinfo{year}{2007}).

\bibitem[{\citenamefont{Odenbach}(2016)}]{odenbach2016microstructure}
\bibinfo{author}{\bibfnamefont{S.}~\bibnamefont{Odenbach}},
  \bibinfo{journal}{Arch. Appl. Mech.} \textbf{\bibinfo{volume}{86}},
  \bibinfo{pages}{269} (\bibinfo{year}{2016}).

\bibitem[{\citenamefont{Weeber et~al.}(2018)\citenamefont{Weeber, Hermes,
  Schmidt, and Holm}}]{weeber2018polymer}
\bibinfo{author}{\bibfnamefont{R.}~\bibnamefont{Weeber}},
  \bibinfo{author}{\bibfnamefont{M.}~\bibnamefont{Hermes}},
  \bibinfo{author}{\bibfnamefont{A.~M.} \bibnamefont{Schmidt}},
  \bibnamefont{and} \bibinfo{author}{\bibfnamefont{C.}~\bibnamefont{Holm}},
  \bibinfo{journal}{J. Phys.: Condens. Matter} \textbf{\bibinfo{volume}{30}},
  \bibinfo{pages}{063002} (\bibinfo{year}{2018}).

\bibitem[{\citenamefont{B{\"o}se et~al.}(2012)\citenamefont{B{\"o}se,
  Rabindranath, and Ehrlich}}]{bose2012soft}
\bibinfo{author}{\bibfnamefont{H.}~\bibnamefont{B{\"o}se}},
  \bibinfo{author}{\bibfnamefont{R.}~\bibnamefont{Rabindranath}},
  \bibnamefont{and} \bibinfo{author}{\bibfnamefont{J.}~\bibnamefont{Ehrlich}},
  \bibinfo{journal}{J. Intel. Mater. Syst. Struct.}
  \textbf{\bibinfo{volume}{23}}, \bibinfo{pages}{989} (\bibinfo{year}{2012}).

\bibitem[{\citenamefont{Hines et~al.}(2017)\citenamefont{Hines, Petersen, Lum,
  and Sitti}}]{hines2017soft}
\bibinfo{author}{\bibfnamefont{L.}~\bibnamefont{Hines}},
  \bibinfo{author}{\bibfnamefont{K.}~\bibnamefont{Petersen}},
  \bibinfo{author}{\bibfnamefont{G.~Z.} \bibnamefont{Lum}}, \bibnamefont{and}
  \bibinfo{author}{\bibfnamefont{M.}~\bibnamefont{Sitti}},
  \bibinfo{journal}{Adv. Mater.} \textbf{\bibinfo{volume}{29}},
  \bibinfo{pages}{1603483} (\bibinfo{year}{2017}).

\bibitem[{\citenamefont{Fischer and Menzel}(2020)}]{fischer2020towards}
\bibinfo{author}{\bibfnamefont{L.}~\bibnamefont{Fischer}} \bibnamefont{and}
  \bibinfo{author}{\bibfnamefont{A.~M.} \bibnamefont{Menzel}},
  \bibinfo{journal}{Phys. Rev. Research} \textbf{\bibinfo{volume}{2}},
  \bibinfo{pages}{023383} (\bibinfo{year}{2020}).

\bibitem[{\citenamefont{Jolly et~al.}(1996)\citenamefont{Jolly, Carlson,
  Mu{\~n}oz, and Bullions}}]{jolly1996magnetoviscoelastic}
\bibinfo{author}{\bibfnamefont{M.~R.} \bibnamefont{Jolly}},
  \bibinfo{author}{\bibfnamefont{J.~D.} \bibnamefont{Carlson}},
  \bibinfo{author}{\bibfnamefont{B.~C.} \bibnamefont{Mu{\~n}oz}},
  \bibnamefont{and} \bibinfo{author}{\bibfnamefont{T.~A.}
  \bibnamefont{Bullions}}, \bibinfo{journal}{J. Intell. Mater. Syst. Struct.}
  \textbf{\bibinfo{volume}{7}}, \bibinfo{pages}{613} (\bibinfo{year}{1996}).

\bibitem[{\citenamefont{Sch{\"u}mann and Odenbach}(2017)}]{schumann2017situ}
\bibinfo{author}{\bibfnamefont{M.}~\bibnamefont{Sch{\"u}mann}}
  \bibnamefont{and} \bibinfo{author}{\bibfnamefont{S.}~\bibnamefont{Odenbach}},
  \bibinfo{journal}{J. Magn. Magn. Mater.} \textbf{\bibinfo{volume}{441}},
  \bibinfo{pages}{88} (\bibinfo{year}{2017}).

\bibitem[{\citenamefont{Temmen et~al.}(2000)\citenamefont{Temmen, Pleiner, Liu,
  and Brand}}]{temmen2000convective}
\bibinfo{author}{\bibfnamefont{H.}~\bibnamefont{Temmen}},
  \bibinfo{author}{\bibfnamefont{H.}~\bibnamefont{Pleiner}},
  \bibinfo{author}{\bibfnamefont{M.}~\bibnamefont{Liu}}, \bibnamefont{and}
  \bibinfo{author}{\bibfnamefont{H.~R.} \bibnamefont{Brand}},
  \bibinfo{journal}{Phys. Rev. Lett.} \textbf{\bibinfo{volume}{84}},
  \bibinfo{pages}{3228} (\bibinfo{year}{2000}).

\bibitem[{\citenamefont{Puljiz and
  Menzel}(2019{\natexlab{a}})}]{puljiz2019memory}
\bibinfo{author}{\bibfnamefont{M.}~\bibnamefont{Puljiz}} \bibnamefont{and}
  \bibinfo{author}{\bibfnamefont{A.~M.} \bibnamefont{Menzel}},
  \bibinfo{journal}{Phys. Rev. E} \textbf{\bibinfo{volume}{99}},
  \bibinfo{pages}{012601} (\bibinfo{year}{2019}{\natexlab{a}}).

\bibitem[{\citenamefont{Dhont}(1996)}]{dhont}
\bibinfo{author}{\bibfnamefont{J.~K.~G.} \bibnamefont{Dhont}},
  \emph{\bibinfo{title}{An Introduction to Dynamics of Colloids}}
  (\bibinfo{publisher}{Elsevier, Amsterdam}, \bibinfo{year}{1996}).

\bibitem[{\citenamefont{Kreissl et~al.}(2021)\citenamefont{Kreissl, Holm, and
  Weeber}}]{kreissl2021frequency}
\bibinfo{author}{\bibfnamefont{P.}~\bibnamefont{Kreissl}},
  \bibinfo{author}{\bibfnamefont{C.}~\bibnamefont{Holm}}, \bibnamefont{and}
  \bibinfo{author}{\bibfnamefont{R.}~\bibnamefont{Weeber}},
  \bibinfo{journal}{Soft Matter} \textbf{\bibinfo{volume}{17}},
  \bibinfo{pages}{174} (\bibinfo{year}{2021}).

\bibitem[{\citenamefont{Ilg and Evangelopoulos}(2018)}]{PhysRevE.97.032610}
\bibinfo{author}{\bibfnamefont{P.}~\bibnamefont{Ilg}} \bibnamefont{and}
  \bibinfo{author}{\bibfnamefont{A.~E. A.~S.} \bibnamefont{Evangelopoulos}},
  \bibinfo{journal}{Phys. Rev. E} \textbf{\bibinfo{volume}{97}},
  \bibinfo{pages}{032610} (\bibinfo{year}{2018}).

\bibitem[{\citenamefont{Kubo et~al.}(1991)\citenamefont{Kubo, Toda, and
  Hashitsume}}]{kubo1991statistical}
\bibinfo{author}{\bibfnamefont{R.}~\bibnamefont{Kubo}},
  \bibinfo{author}{\bibfnamefont{M.}~\bibnamefont{Toda}}, \bibnamefont{and}
  \bibinfo{author}{\bibfnamefont{N.}~\bibnamefont{Hashitsume}},
  \emph{\bibinfo{title}{Statistical Physics II: Nonequilibrium Statistical
  Mechanics}} (\bibinfo{publisher}{Springer, Berlin}, \bibinfo{year}{1991}).

\bibitem[{\citenamefont{Zwanzig}(2001)}]{zwanzig2001nonequilibrium}
\bibinfo{author}{\bibfnamefont{R.}~\bibnamefont{Zwanzig}},
  \emph{\bibinfo{title}{Nonequilibrium Statistical Mechanics}}
  (\bibinfo{publisher}{Oxford University Press, Oxford}, \bibinfo{year}{2001}).

\bibitem[{\citenamefont{Phan-Thien}(1993)}]{phan1993rigid}
\bibinfo{author}{\bibfnamefont{N.}~\bibnamefont{Phan-Thien}},
  \bibinfo{journal}{J. Elasticity} \textbf{\bibinfo{volume}{32}},
  \bibinfo{pages}{243} (\bibinfo{year}{1993}).

\bibitem[{\citenamefont{Phan-Thien and Kim}(1994)}]{phan1994load}
\bibinfo{author}{\bibfnamefont{N.}~\bibnamefont{Phan-Thien}} \bibnamefont{and}
  \bibinfo{author}{\bibfnamefont{S.}~\bibnamefont{Kim}}, \bibinfo{journal}{Z.
  Angew. Math. Phys.} \textbf{\bibinfo{volume}{45}}, \bibinfo{pages}{177}
  (\bibinfo{year}{1994}).

\bibitem[{\citenamefont{Puljiz and Menzel}(2017)}]{puljiz2017forces}
\bibinfo{author}{\bibfnamefont{M.}~\bibnamefont{Puljiz}} \bibnamefont{and}
  \bibinfo{author}{\bibfnamefont{A.~M.} \bibnamefont{Menzel}},
  \bibinfo{journal}{Phys. Rev. E} \textbf{\bibinfo{volume}{95}},
  \bibinfo{pages}{053002} (\bibinfo{year}{2017}).

\bibitem[{\citenamefont{Puljiz and
  Menzel}(2019{\natexlab{b}})}]{puljiz2019displacement}
\bibinfo{author}{\bibfnamefont{M.}~\bibnamefont{Puljiz}} \bibnamefont{and}
  \bibinfo{author}{\bibfnamefont{A.~M.} \bibnamefont{Menzel}},
  \bibinfo{journal}{Phys. Rev. E} \textbf{\bibinfo{volume}{99}},
  \bibinfo{pages}{053002} (\bibinfo{year}{2019}{\natexlab{b}}).

\bibitem[{\citenamefont{Strobl}(2007)}]{strobl2007physics}
\bibinfo{author}{\bibfnamefont{G.}~\bibnamefont{Strobl}},
  \emph{\bibinfo{title}{The Physics of Polymers}}
  (\bibinfo{publisher}{Springer, Berlin}, \bibinfo{year}{2007}).

\bibitem[{\citenamefont{Narinder et~al.}(2019)\citenamefont{Narinder,
  Gomez-Solano, and Bechinger}}]{narinder2019active}
\bibinfo{author}{\bibnamefont{Narinder}}, \bibinfo{author}{\bibfnamefont{J.~R.}
  \bibnamefont{Gomez-Solano}}, \bibnamefont{and}
  \bibinfo{author}{\bibfnamefont{C.}~\bibnamefont{Bechinger}},
  \bibinfo{journal}{New J. Phys.} \textbf{\bibinfo{volume}{21}},
  \bibinfo{pages}{093058} (\bibinfo{year}{2019}).

\bibitem[{\citenamefont{Qi et~al.}(2020)\citenamefont{Qi, Westphal, Gompper,
  and Winkler}}]{qi2020enhanced}
\bibinfo{author}{\bibfnamefont{K.}~\bibnamefont{Qi}},
  \bibinfo{author}{\bibfnamefont{E.}~\bibnamefont{Westphal}},
  \bibinfo{author}{\bibfnamefont{G.}~\bibnamefont{Gompper}}, \bibnamefont{and}
  \bibinfo{author}{\bibfnamefont{R.~G.} \bibnamefont{Winkler}},
  \bibinfo{journal}{Phys. Rev. Lett.} \textbf{\bibinfo{volume}{124}},
  \bibinfo{pages}{068001} (\bibinfo{year}{2020}).

\bibitem[{\citenamefont{Zöttl and Yeomans}(2019)}]{Zottl.2019}
\bibinfo{author}{\bibfnamefont{A.}~\bibnamefont{Zöttl}} \bibnamefont{and}
  \bibinfo{author}{\bibfnamefont{J.~M.} \bibnamefont{Yeomans}},
  \bibinfo{journal}{Nature Physics} \textbf{\bibinfo{volume}{15}},
  \bibinfo{pages}{554} (\bibinfo{year}{2019}), ISSN \bibinfo{issn}{1745-2481}.

\bibitem[{\citenamefont{Martinez-Pedrero
  et~al.}(2018)\citenamefont{Martinez-Pedrero, Navarro-Argem{\'\i},
  Ortiz-Ambriz, Pagonabarraga, and Tierno}}]{martinez2018emergent}
\bibinfo{author}{\bibfnamefont{F.}~\bibnamefont{Martinez-Pedrero}},
  \bibinfo{author}{\bibfnamefont{E.}~\bibnamefont{Navarro-Argem{\'\i}}},
  \bibinfo{author}{\bibfnamefont{A.}~\bibnamefont{Ortiz-Ambriz}},
  \bibinfo{author}{\bibfnamefont{I.}~\bibnamefont{Pagonabarraga}},
  \bibnamefont{and} \bibinfo{author}{\bibfnamefont{P.}~\bibnamefont{Tierno}},
  \bibinfo{journal}{Science Adv.} \textbf{\bibinfo{volume}{4}},
  \bibinfo{pages}{eaap9379} (\bibinfo{year}{2018}).

\end{thebibliography}

\end{document}